# Structure and dynamics of DOPC vesicles: A transformation from unilamellar to multilamellar vesicles by n-alkyl-PEO polymer

Judith U. De Mel,[1] Sudipta Gupta,[1] Lutz Wilner,[2] Jürgen Allgaier,[2] Laura R. Stingaciu,[3] Markus Bleuel,[4] and Gerald J. Schneider[1,5]

[1]*Department of Chemistry, Louisiana State University, Baton Rouge, LA 70803, USA*

[2] *Jülich Center for Neutron Science (JCNS-1) and Institute of Biological Information Processing (IBI-8) Forschungszentrum Jülich GmbH, 52428 Jülich, Germany*

[3]*Neutron Sciences Directorate, Oak Ridge National Laboratory (ORNL), POB 2008, 1 Bethel Valley Road, TN 37831, Oak Ridge, USA*

[4]*NIST Center for Neutron Research, National Institute of Standards and Technology, Gaithersburg, Maryland 20899-8562, USA*

[5]*Department of Physics & Astronomy, Louisiana State University, Baton Rouge, LA 70803, USA*

## ABSTRACT

We investigate the influence of a non-ionic surfactant like polymer on phospholipid vesicles. Our results from cryogenic transmission electron microscopy (cryo-TEM), dynamic light scattering (DLS), small angle neutron and X-ray scattering (SANS/SAXS), identifies the existence of multilayer vesicles and an increase in size of the vesicles in presence of the polymers. We present a generalized model to obtain the bending rigidity from neutron spin echo spectroscopy (NSE) data for multilayer vesicles. We demonstrated that polymers are trapped in the lipid bilayer, causing a partial disruption in the vesicle, which is attributed to the reduction in bending rigidity per unit bilayer. We also observed substantial dampening of the trapped lipid tail motion in presence of the polymer. Our results highlighted the possibilities of using specialized polymers that can disrupt membrane and control their dynamics with possible application in topical drug or nutraceutical formulations.

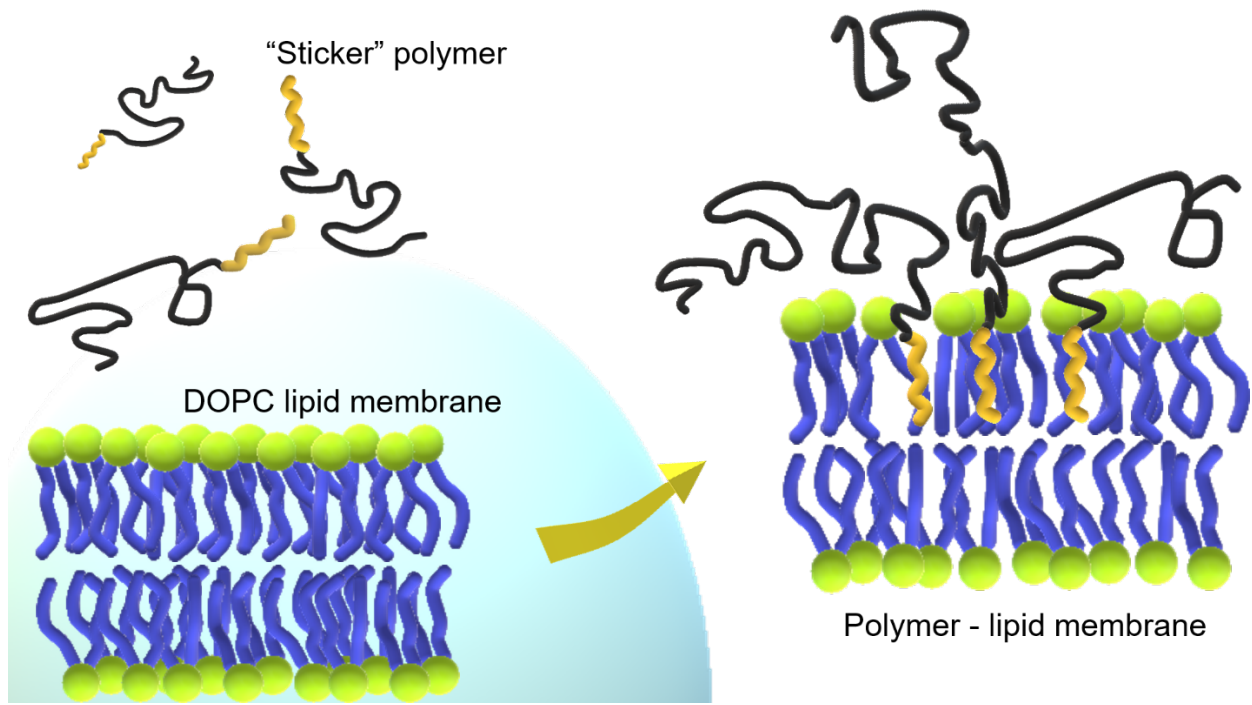

# 1   INTRODUCTION

Phospholipid vesicles are a versatile system for a variety of applications.[1-3] Primarily made up of phospholipids which have a polar head group and hydrophobic lipid tail/tails, these are spherical self-assemblies that can span in sizes from tens of nanometers to several micrometers. The general structure is universal for vesicles where it consists of a solution core encapsulated by a phospholipid bilayer with the polar head groups of the two layers facing in opposite directions. The interior solution core, and the outside bulk solution allowing the lipid tails to be sandwiched in the middle. Properties of phospholipid vesicles can be changed using additives such as cholesterol, small molecules, and macromolecules such as synthetic polymers.[4-6,7,8,9,10,11, 12]  From the initial stages of incorporating synthetic polymers to phospholipid vesicles, the present vesicle formulations with associated polymers have come a long way in terms of fine-tuning their properties for drug delivery and applications in cosmetics, nutraceuticals, food technology etc.[13]

Mainly, polymer-liposome interactions could lead to i) modifications (coating, insertion) ii) disruptions or iii) transformation of self-assembly of the lipid bilayer structure.[14] These interactions are determined by the liposome morphology, bilayer structure and dynamics in presence of polymers, as well as external physical parameters, such as solvent quality, temperature, ionic strength, pressure etc. Altogether, they can unfold a plethora of benefits in applications when it comes to i) encapsulation efficiency, ii) controlled release or iii) drug transport mechanisms such as skin penetration.[15, 16] Despite the abundance of applications pointing towards the importance of molecular level interactions, fundamental questions such as how polymer-liposome interactions at molecular level drive structure and dynamic changes are not well understood.

One well-known application of synthetic polymers with phospholipid membranes is the use of PEG (polyethylene glycol) polymers. PEG has the basic structure similar to PEO (polyethylene oxide) or POE (Polyolefin), -(CH$_2$-CH$_2$-O-)-. They are used as steric stabilization agents to formulate "stealth" vesicles with increased blood circulation time.[17] Depending on the area density of PEO chains equipped with hydrophobic anchors, like cholesterol or a pair of dodecyl groups can attach onto the outer lipid layer, while the chains can exhibit a "mushroom" regime or "brush" regime and change overall rigidity of the bilayer.[18] PEG when is interacted with proteins is known to show various levels of cytotoxicity and aggregations.[19, 20] Another notable example is PEO - block copolymers. The PEO - PPO (polypropylene oxide) di-block or PEO - PPO - PEO tri-block copolymers also known as polaxomers have emerged as an alternative to PEG-grafted lipids and are used extensively.[8, 9] An intermediate between these two cases could be polyoxyethylene alkyl ethers commonly known as Brij surfactants. Polyoxyethylene alkyl ethers (C$_i$E$_j$) consisting a hydrophilic polyethylene oxide chain and a hydrophobic alkyl chain are non-ionic surfactants that can show multiple self-assembly structures.[21-23] These are commonly used in soaps and detergents in combination with other surfactants. Recently, they have also caught attention in biomedical applications.[24, 25] Polymers with C$_i$E$_j$ type structures are extensively used in cosmetics, personal care and consumer products as emulsifiers and surfactants. The non-ionic nature combined with the amphiphilic nature allows these polymers to show unique properties depending on the environmental conditions such as hydrophilic-hydrophilic balance (HLB), pH, temperature, ionic strength, lipid head group.[26]

In previous studies it has been shown that non-ionic surfactants can modify lipid bilayers in numerous ways.[27-29] Detergents such as Triton x-100 have shown to modify the ion permeability of lipid membranes.[30] Similarly, $C_iE_j$ polymers which belong to the class of non-ionic surfactants can cause several modifications of phospholipid bilayers. Work of Liu *et. al.* has shown that $C_iE_j$ polymers can insert into phospholipid bilayers by the hydrophobic end using quartz crystal microbalance (QCM) experiments.[10] The chain length of the hydrophobic segment is shown to play a key role in determine the successful insertion to the bilayer.[8, 31] In the work by Gutberlet *et al.* using POPC and $C_{12}EO_n$ (n = 6,8) systems they have shown the molar ratio of polymeric surfactant can decrease the thickness of bilayer and gradually transform the system to different micellar structures.[11]

This study investigates the structural and dynamic changes of the lipid membrane caused by the incorporation of the polymeric surfactant poly(ethylene oxide)-mono-n-octadecylether, $C_{18}$-PEO4 on DOPC (1,2-dioleoyl-sn-glycero-3-phosphocholine). The $C_{18}$-PEO4, consists of an n-octadecyl hydrophobic alkyl tail, and a ~ 4 kg/mol hydrophilic PEO chain. Similar polymers have been shown to play a key role, in the interaction with membrane for integral proteins,[32, 33] as well as slowing down lipid and vitamin-E oxidation.[34] We wanted to determine the effect of $C_{18}$-PEO4 polymers in membrane penetration and dynamics to obtain useful properties for topical drug or nutraceutical formulations.

## 2    THEORETICAL BACKGROUND

In this section we present the data modelling theory used to understand the macroscopic scattering cross-section, $d\Sigma/d\Omega$, for the vesicle structure and the membrane. Both the SANS and SAXS experiments are performed at ambient temperature of 20 °C, which corresponds to the fluid phase of DOPC.[35]

### 2.1    Vesicle structure

A modified core-shell model is used to describe the vesicle form factor.[36, 37] As illustrated in Figure 1 the core is filled with water which is encapsulated by $N$ shells of lipids and $N$-1 layers of solvent in case of multilayer vesicles (MLVs). The thickness and scattering length density of each shell is assumed to be identical. The corresponding form factor is given by:

$$P(Q, R, t, \Delta\rho) = \frac{\phi[F(Q)]^2}{V(R_N)} \qquad (1)$$

with

$$F(Q) = (\rho_{\text{shell}}$$
$$- \rho_{\text{solv}}) \sum_{i=1}^{N} \left[ 3V(r_i) \frac{\sin(Qr_i) - Qr_i \cos(Qr_i)}{(Qr_i)^3} \right. \qquad (2)$$
$$\left. - 3V(R_i) \frac{\sin(QR_i) - Qr_i \cos(QR_i)}{(QR_i)^3} \right]$$

For

$$r_i = r_c + (i-1)(t_s + t_w) \qquad \text{solvent radius before shell } i \qquad (3)$$
$$R_i = r_i + t_s \qquad\qquad\qquad \text{shell radius for shell } i$$

Here, $V(r)$ is the volume of the sphere with radius $r$, $r_c$ is the radius of the core, $t_s$ is the thickness of the individual shells, $t_w$ is the thickness of the inter-bilayer water, $\phi$, the corresponding lipid volume fraction. The outer perimeter radius is given by, $R_{SANS} = r_c + Nt_s + (N-1)t_w$. For DOPC we used the scattering length density of the shell, $\rho_{\text{shell}} = 3.01 \times 10^9$ cm$^{-2}$ and for D$_2$O the scattering length density of the solvent, $\rho_{\text{solv}} = 6.36 \times 10^{10}$ cm$^{-2}$, respectively.[38] The macroscopic scattering cross-section is obtained from the scattering intensity in SANS and is given by

$$\frac{d\Sigma}{d\Omega}(Q)_{SANS} = I(Q)_{SANS} = \int dr P(Q, R, t, \rho_{\text{lipo}}, \rho_{\text{solv}}) s(r) \qquad (4)$$

For the size polydispersity, $s(r)$, we used a Schulz distribution and a log-normal distribution. In addition, the thickness of the shell and the solvent are convoluted with a Gaussian distribution function to account for the thickness polydispersity.

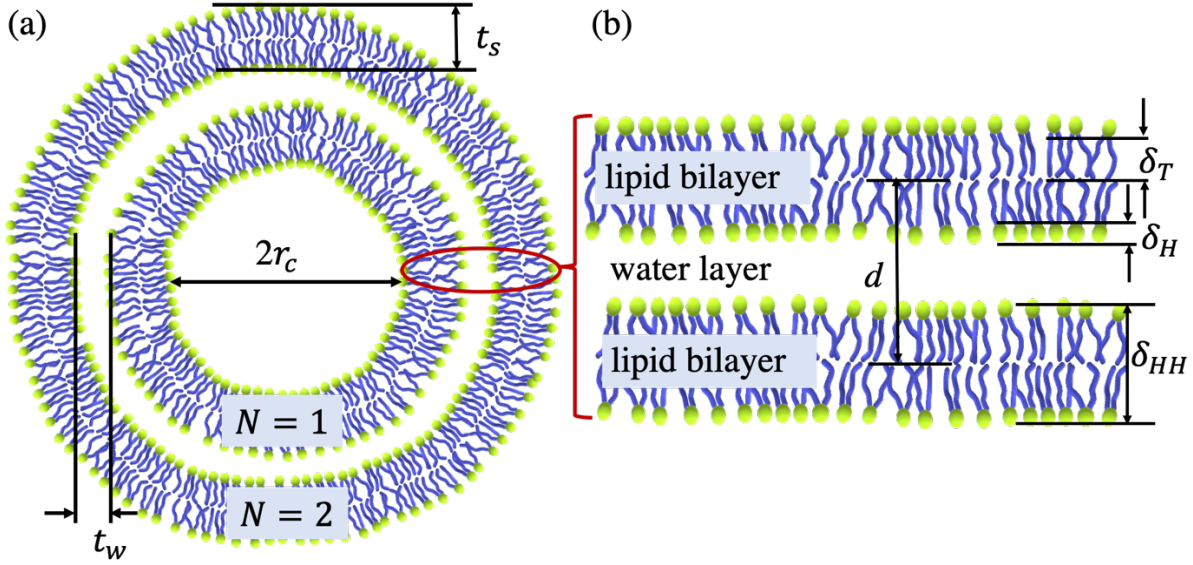

*Figure 1. Schematic representation of (a) the multilayer liposome, and (b) lipid multilayers illustrating the number of bilayers, N, the radius of the core, $r_c$, the thickness of the individual shells, $t_s$, the thickness of the interleaved solvent layers, $t_w$, the thickness of the lipid head, $\delta_H$, the thickness of the lipid tail region, $\delta_T$ and the lamellar repeat distance, d, of bilayers.*

## 2.2   Hard sphere structure

The scattering intensity from non-interacting uniform hard spheres (HS) is given by the product of the form factor and the static structure factor. It is given by:[39]

$$I(Q)_{HS} = \int dr P(Q)_{HS} S(Q)_{HS} s(r) \tag{5}$$

Where the HS form factor is given by,[39]

$$P(Q)_{HS} = \frac{\phi_{HS}}{V} \Delta\rho^2 \left[ 3 \frac{\sin(QR_{HS}) - (QR_{HS})\cos(QR_{HS})}{(QR_{HS})^3} \right]^2 \tag{6}$$

With, V, the volume of the HS nanoparticle (NP) with $\phi_{HS}$, the corresponding HS volume fraction. $R_{HS}$ is the radius, and $\Delta\rho = \rho_{\text{lipo}} - \rho_{solv}$ is the contrast difference of the NP with respect to the solvent. The scattering from inter-particle contribution is given the HS structure factor, $S(Q)_{HS}$, given by the solution of Ornstein-Zernike equation using Percus-Yevick approximation.[39] The size polydispersity in equation 5 is given by $s(r)$, similar to equation 4.

## 2.3 Membrane structure

The random lamellar sheet consisting of the heads and tails of the phospholipids can be modelled using the Caille structure factor.[40, 41] SAXS provides direct access to the macroscopic scattering cross-section given by the scattering intensity, and for a random distribution of the lamellar phase it is given by

$$\frac{d\Sigma}{d\Omega}(Q)_{SAXS} = I(Q)_{SAXS} = 2\pi \frac{VP(Q)S(Q)}{Q^2 d} \tag{7}$$

with the scattering volume, $V$, and the distance of the lamellae, $d$. The form factor is given by:

$$P(Q) = \frac{4}{Q^2}\left[\Delta\rho_{\text{H}}\left\{\sin\left(Q(\delta_H + \delta_T)\right) - \sin(Q\delta_T)\right\} + \Delta\rho_{\text{T}}\sin(Q\delta_T)\right]^2 \tag{8}$$

The scattering contrasts for the head and tail are $\Delta\rho_{\text{H}}$ and $\Delta\rho_{\text{T}}$, respectively. The corresponding thicknesses are $\delta_H$ and $\delta_T$, respectively, as presented in Figure 1. The head to head bilayer thickness is given by, $\delta_{HH} = 2(\delta_H + \delta_T)$. The Caille structure factor is given by

$$S(Q) = 1 + 2\sum_{n=1}^{N-1}\left(1 - \frac{n}{N}\right)\cos(Qdn)\exp\left(-\frac{2Q^2 d^2 \alpha(n)}{2}\right) \tag{9}$$

with the number of lamellar plates, $N$, and the correlation function for the lamellae, $\alpha(n)$, defined by

$$\alpha(n) = \frac{\eta_{\text{cp}}}{4\pi^2}(\ln(\pi n) + \gamma_E) \tag{10}$$

with $\gamma_E = 0.57721$ the Euler's constant. The elastic constant for the membranes is expressed in terms of the Caille parameter, $\eta_{\text{cp}} = \frac{Q_0^2 k_B T}{8\pi\sqrt{\kappa_b \, \kappa_A / \delta_{HH}}}$, where $\kappa_b$ and $\kappa_A$ are the bending elasticity and the compression modulus of the membranes, respectively. Here $\kappa_A$ is associated with the interactions between the membranes. The position of the first-order Bragg peak is given by $Q_0$, and $k_B$ is the Boltzmann's constant and $T$ the absolute temperature.

## 2.4 Vesicle dynamics

Neutron spin echo (NSE) spectroscopy has proven to be a powerful tool to follow the molecular motions in vesicles.[42-44] This method reaches the highest energy resolution ($\sim$ neV) of all available neutron scattering spectrometers and therefore allows to measure the dynamic structure factor or the intermediate scattering function, $S(Q,t)$, up to several hundred nanoseconds.

Recently, it has been shown that diffusion, membrane fluctuations, and confined motion of lipid tails lead to major contributions for the modeling of the intermediate scattering function.[45] The statistically independent tail-motion and height-height correlation resulting in membrane undulations can be coupled to the overall translational diffusion of the liposome as:[45]

$$S_{liposome}(Q,t) = \left( n_{H,head} + n_{H,tail} \left( \mathcal{A}(Q) + \left( 1 - \mathcal{A}(Q) \right) \exp\left( -\left( \frac{t}{\tau} \right)^\beta \right) \right) \right) S_{ZG}(Q,t) \exp(-D_t Q^2 t) \tag{11}$$

The relative fractions of protons in the head is kept fixed to, $n_{H,head} = 0.21$, for h-DOPC, and, $n_{H,tail} = 1 - n_{H,head} = 0.79$. The average scattering length density of the lipid head or tail gets smaller with proton density. The variable $\mathcal{A}(Q)$ refers to the elastic fraction of the lipid tail motion and can be compared with the elastic incoherent structure factor (EISF) from quasielastic neutron scattering (QENS).[45] The first part signifies the motion of the lipid molecule, the second and third part reflects the membrane undulations. The membrane undulation can be well described by the Zilman-Granek (ZG) approach given by a stretched-exponential decay:[46]

$$S_{ZG}(Q,t) = A \exp\left[ -\left( \Gamma_Q t \right)^{2/3} \right] \tag{12}$$

The free parameters are the $Q$-dependent decay rate, $\Gamma_Q$, and the amplitude, $A$.

The effective bending modulus, $\tilde{\kappa}$ is calculated from the modified ZG theory by Watson and Brown[47] as:

$$\frac{\Gamma_q}{Q^3} = 0.025 \gamma \frac{k_B T}{\eta} \sqrt{\frac{k_B T}{\tilde{\kappa}}} \tag{13}$$

Unlike membranes in bicontinuous microemulsion[48] the effective bending modulus in lipid membrane, $\frac{\tilde{\kappa}}{k_B T} \gg 1$, therefore, $\gamma = 1$ is a reasonable approximation as suggested by Zilman and Granek.[46] Here, $\eta$ is the solvent viscosity, $k_B$, the Boltzmann constant and, $T$, the temperature in absolute scale.

The effective bending modulus, $\tilde{\kappa}$, (or dynamic curvature modulus) is related to the bilayer curvature modulus, $\kappa_\eta$, given by $\tilde{\kappa} = \kappa_\eta + 2h^2 k_m$.[47] Here $\kappa_\eta$ is the parameter of interest which can be obtained from NSE. The monolayer area compressibility modulus for uniform plates of monolayers can be related to the monolayer bending rigidity, $\kappa_b$, as, $k_m = 12\kappa_b/h_t^2$.[49] Here, $h_t$ is thickness of the tail only region of the membrane (monolayer hydrocarbon thickness), and $h$ is the monolayer thickness or the height of the neutral surface from the bilayer midplane given by $h = \delta_{HH}/2$. To express the monolayer parameter, $\kappa_m$, in terms of bilayer parameter, $\kappa_\eta$, we can use, $\kappa_\eta = 2\kappa_b$, and $\tilde{\kappa}$ can be expressed for a bilayer as, $\tilde{\kappa} = \kappa_\eta\{1 + 48(h/2h_t)^2\}$.[50] Considering the neutral surface as the interface between hydrophilic head group and the hydrophobic tail ($h = h_t$),[51-55] we can redefine equation (6) to obtain the bending rigidity of a bilayer from ULV [50]

$$\Gamma_q = 0.0069 \frac{k_B T}{\eta} \sqrt{\frac{k_B T}{\kappa_\eta}} \tag{14}$$

Equation 14 has been successfully used to calculate $\kappa_\eta$ for ULVs from NSE.[37, 50, 56]

For the special case of 4 monolayers, i.e. 2 bilayers or $N = 2$ (neglecting any elastic effects from inter-bilayer water since they are predominantly viscous) we can redefine $\kappa_\eta = 4\kappa_b$ and $\tilde{\kappa} = \kappa_\eta\{1 + 24(h/2h_t)^2\}$, which results in

$$\frac{\Gamma_q}{Q^3} = 0.0094 \frac{k_B T}{\eta} \sqrt{\frac{k_B T}{\kappa_\eta}} \tag{15}$$

In general, for $N$ layers i.e. $2N$ monolayers we have $\kappa_\eta = 2N\kappa_b$ and $\tilde{\kappa} = \kappa_\eta(1 + 12/N)$, which results in

$$\frac{\Gamma_q}{Q^3} = 0.025 \frac{k_B T}{\eta} \sqrt{\frac{k_B T}{\kappa_\eta(1 + 12/N)}} \tag{16}$$

According to this expression the effective bending modulus for MLVs increases by a factor $\frac{13N}{(N+12)}$ for $N \geq 2$.

Additionally, we can analyze the mean squared displacement ($\langle \Delta r(t)^2 \rangle$, MSD) and the non-Gaussianity parameter, $\alpha_2(t) = \frac{d}{d+2} \frac{\langle \Delta r(t)^4 \rangle}{\langle \Delta r(t)^2 \rangle^2} - 1$, from the measured dynamic structure factor, $S(Q,t)$, using a cumulant expansion given by, [37, 45, 57, 58]

$$\frac{S(Q,t)}{S(Q)} = A \exp\left[ -\frac{Q^2 \langle \Delta r(t)^2 \rangle}{6} + \frac{Q^4 \alpha_2(t)}{72} \langle \Delta r(t)^2 \rangle^2 \right] \qquad (17)$$

The non-Gaussianity parameter $\alpha_2$ is essentially defined as quotient of the fourth $\langle \Delta r(t)^4 \rangle$ and the second moment squared $\langle \Delta r(t)^2 \rangle^2$ and $d = 3$, is the dimension of space.[37, 58, 59] Following equations 12, 14 and 17 we can express the membrane rigidity for ULVs as a function of Fourier time, given by[45]

$$\frac{\kappa_\eta}{k_B T} = \frac{t^2}{c(\eta, T)^3 \langle \Delta r(t)^2 \rangle^3} \qquad (18)$$

with $c(\eta, T) = \frac{1}{6}\left( \frac{\eta}{0.0069 k_B T} \right)^{2/3}$. For ZG approximation $\langle \Delta r(t)^2 \rangle \propto t^{2/3}$ and the bending rigidity as a function of time should yield, $\kappa_\eta / k_B T \propto t^2/t^2 = $ constant. Any deviation from this constant behavior will reflect additional dynamics that is not considered by the ZG model. For MLVs the prefactor $c(\eta, T)$ needs to be modified following equation 16.

# 3 MATERIALS AND METHODS

## 3.1 Sample preparation

All chemicals and reagents were used as received. 1,2-di-(octadecenoyl)-*sn*-glycero-3-phosphocholine (DOPC) was purchased from Avanti Polar Lipids (Alabaster, AL, USA), and $D_2O$ were received from Sigma Aldrich (St. Louis, MO, USA). We have used fully deuterated poly-(ethylene oxide)-mono-n-octadecylether, $dC_{18}$-PEO4, to reduce scattering contributions resulting from contrast between solvent and polymer. The polymer was synthesized by living ring-opening, anionic polymerization of fully deuterated ethylene oxide, d-EO. The initiator was a mixture of deuterated 1-octacosanol and the corresponding potassium 1-octacosanolate. Exact polymerization conditions can be found in reference [60]. The polymer was characterized by size exclusion chromatography using a combination of refractive index and 18-angle light scattering detector (Optilab rEX and DAWN HELEOS-II, Wyatt) for absolute molecular weight characterization. For separation three Agilent Plus Pore GPC columns with continuous pore size distribution were taken and a mixture of tetrahydrofuran, dimethylacetamid, and acetic acid as eluent at a flux rate of 1mL/min. The degree of polymerisation of dPEO is 92 and the dispersity index 1.03. This polymer is considered to be hybrid between block copolymer and non-ionic surfactant.[61]

DOPC vesicles were prepared by dissolving DOPC lipid powder in chloroform and removing the solvent using a rotary evaporator and drying further under vacuum overnight. The dried lipid was hydrated using $D_2O$ and the resultant solution was subjected to freeze-thaw cycling by alternatingly immersing the flask in water at around 50 °C and placing in a freezer at -20 °C in ten-minute intervals. Finally, the solution was extruded using a mini extruder (Avanti Polar Lipids, Alabaster, AL, USA) through a polycarbonate membrane with pore diameter of 100 nm (33 passes) to obtain unilamellar vesicles. Vesicle solutions were mixed with $dC_{18}$-PEO4 solutions to obtain the desired external polymeric concentrations. This technique ensures the addition of the polymer form outside the vesicles. Measurements for each mixture were averaged starting 24 hours after sample preparation. All experiments were conducted at ambient temperature.

## 3.2   DLS

Dynamic light scattering (DLS) measurements were performed using a Malvern Zetasizer Nano ZS equipped with a He-Ne laser of wavelength, $\lambda = 633$ nm at 30 mW laser power, at a scattering angle $\theta = 173°$. The hydrodynamic radius, $R_h$, of the liposomes in each $dC_{18}$-PEO4 concentration was calculated from the translational diffusion coefficient, $D_t$, using the Stokes-Einstein equation, $R_h = k_B T/(6\pi\eta_0 D_t)$, with the Boltzmann constant, $k_B$, the temperature, $T$, the viscosity of the solvent, $\eta_0$. Four separate DLS measurements for each mixture were averaged. In Table 1 the results from DLS are reported for 5% DOPC along with three different concentrations of $dC_{18}$-PEO4 polymer mixed with 5% DOPC. To calculate the corresponding $R_h$ we have used the viscosity of 5% DOPC as the solvent viscosity, $\eta_0 = (1.6 \pm 0.03) \times 10^{-3}$ Pa·s (cf. supplementary information).

## 3.3   Cryo-TEM

Cryogenic-transmission electron microscopy (cryo-TEM) images were recorded on a Tecnai G2 F30 operated at 150 kV. A volume of ten microliters of the sample (0.125 wt% DOPC: in pure $D_2O$, or NaCl) was applied to a 200-mesh lacey carbon grid mounted on the plunging station of a FEI Vitrobot™ and excess liquid was blotted for 2 s by the filter paper attached to the arms of the Vitrobot. The carbon grids with the attached thin film of liposome suspensions were plunged into liquid ethane and transferred to a single tilt cryo - specimen holder for imaging. By quick plunging into liquid ethane, the vesicles are preserved at their hydrated state present at room temperature. Cryo-TEM images were obtained in the bright field setting.

## 3.4 SANS

Small-angle neutron scattering (SANS) experiments were conducted at the NG 7 SANS instrument of the NIST Center for Neutron Research (NCNR) at National Institute of Standards and Technology (NIST).[62] The sample-to-detector distances, $d$, were set to 1, 4, and 13 m, at neutron wavelength, $\lambda = 6$ Å. Another configuration with lenses at $d = 15.3$ m, and $\lambda = 8$ Å was used to access low $Q$'s.[63] This combination covers a $Q$ - range from $\approx 0.001$ to $\approx 0.6$ Å$^{-1}$, where $Q = 4\pi \sin(\theta/2)/\lambda$, with the scattering angle, $\theta$. A wavelength resolution of, $\Delta\lambda/\lambda = 14\%$, was used. All data reduction into intensity, $I(Q)$, vs. momentum transfer, $Q = |\vec{Q}|$, was carried out following the standard procedures that are implemented in the NCNR macros for the Igor software package.[64] The intensity values were scaled into absolute units (cm$^{-1}$) using direct beam. The $D_2O$ as solvent and empty cell were measured separately. The polymer solution composed of 1% dC$_{18}$-PEO4 polymer in $D_2O$ was measured separately and subtracted as background.

## 3.5 SAXS

Small-angle X-ray scattering (SAXS) experiments were conducted at the LIX beamline at the National Synchrotron Light Source II, Brookhaven National Laboratory and at the Bio-SAXS beamline at the Stanford Linear Accelerator Center (SLAC) facility. At the synchrotron instrument, the samples were measured in a flow cell with an acquisition time of 1 s, whereas the samples were loaded in 1 mm borosilicate glass capillary cylinders for the lab X-ray with an acquisition time of 10 s. The recorded intensities were corrected for dark current, empty cell and solvent (buffer) using standard procedures.[65] [39] The polymer solution composed of 1% dC$_{18}$-PEO4 polymer in $D_2O$ was measured separately and subtracted as background.

## 3.6 NSE

Neutron spin echo spectroscopy (NSE) spectroscopy has been used to examine the effects of membrane dynamics simultaneously over broad length and time scales. We obtained NSE data at BL15 at the Spallation Neutron Source of the Oak Ridge National Laboratory, Oak Ridge, TN.[66] We used Hellma quartz cells at BL15-NSE with 2 mm sample thickness. Lipid concentration was always 5%. The data reduction was performed with the standard ECHODET software package of the SNS-NSE instrument. Wavelengths of 8 Å was used at BL15-NSE. $D_2O$ mixed with 1% $dC_{18}$-PEO4 polymer was measured separately and subtracted as background.

# 4    RESULTS AND DISCUSSIONS

## 4.1    Structure and Morphology

The liposome size is determined by analyzing the intensity autocorrelation functions, $g^2(Q,t)$, from DLS using a single exponential decay:

$$g^2(Q,t) = A \exp(-2D_t Q^2 t) \tag{19}$$

The corresponding diffusion coefficients and hydrodynamic radii are summarized in Table 1 for different concentrations of $dC_{18}$-PEO4 polymers in 5 % DOPC. The corresponding diffusion distribution is reported in the supplemental information.

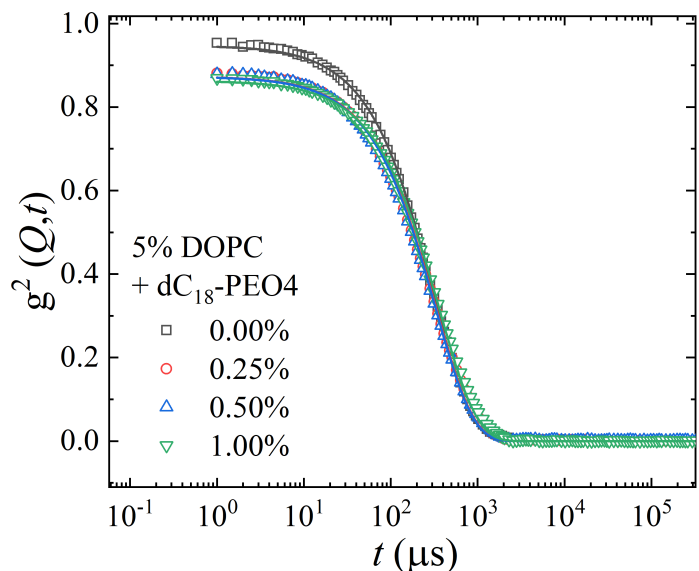

*Figure 2. Intensity autocorrelation function of DOPC 5wt% vesicles with varying $dC_{18}$-PEO4 polymer concentrations. The solid lines represent the single diffusion fit (equation 19). Results are tabulated in Table 1.*

*Table 1. Dynamic light scattering results of DOPC 5 wt% vesicles with varying dC$_{18}$-PEO4 polymer concentration - log normal size distribution, diffusion coefficient, D$_t$, and hydrodynamic radius, R$_h$, analysis.*

| Samples with 5wt% DOPC + dC$_{18}$-PEO4 | Log normal size distribution (fit) % | $D_t \times 10^{-12}$ (m$^2$s$^{-1}$) | $R_h$ (nm) |
|---|---|---|---|
| 0 wt% | 50.2 ± 0.5 | 2.29 ± 0.01 | 60 ± 2 |
| 0.25 wt% | 48.2 ± 0.5 | 2.17 ± 0.01 | 62 ± 2 |
| 0.5 wt% | 49.8 ± 0.5 | 2.22 ± 0.01 | 61 ± 2 |
| 1 wt% | 49.5 ± 0.5 | 1.98 ± 0.01 | 68 ± 2 |

Figure 3 (a) illustrates the cryo-TEM images for 0.25 % mass fraction of DOPC dispersed in 1 % dC$_{18}$-PEO4 polymers. It shows a presence of highly polydisperse mixture of unilamellar and multilayer vesicles (MLV). The data analysis yields a log-normal size distribution for the vesicle size, the core size and the intermembrane distance as shown in Figure 3 (b), (c) and (d), respectively. The corresponding vesicle radius yields, $R_{TEM}$ = 52 ± 4 nm, with 34 ± 9% polydispersity. The water core size analysis yields a core radius, $R_{c,TEM}$ = 28 ± 1 nm, with 52 ± 5 % polydispersity. The intermembrane distance in Figure 3 (c) depicts the average distance between the individual layers inside a single MLV, and is given by $t_{w,TEM}$ = 14.4 ± 0.7 nm, with 39 ± 4% polydispersity. We obtained $R_{TEM}$ , $R_{c,TEM}$, and $t_{w,TEM}$ by counting as many as 187, 106 and about 278 distances, respectively including different orientations. Therefore, these numbers do not reflect the statistical significance of the particle sizes, which will be examined in details from scattering experiments. The results are reported in Table 2. The size analysis for pure DOPC is presented in the SI based on our earlier data.[37] A significant reduction in the size of the water core of DOPC vesicles is observed in presence of dC$_{18}$-PEO4 polymers.

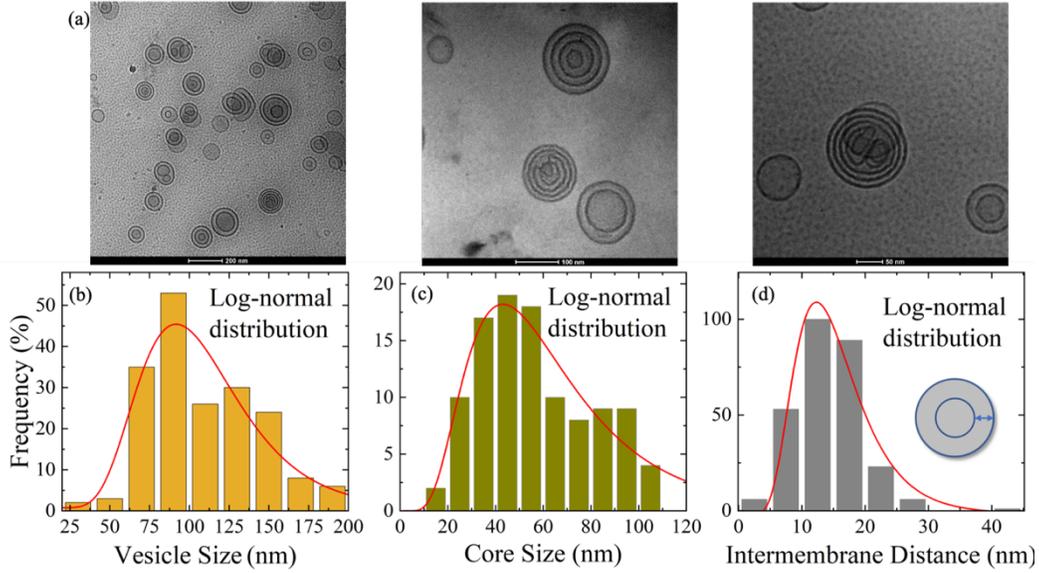

*Figure 3. (a) Cryo-TEM images for 0.25 % mass fraction of DOPC dispersed in 1 % dC₁₈-PEO4 polymers. Corresponding log-normal size distribution of (b) the vesicle size, (c) The water core size and (c) the intermembrane distance, as illustrated schematically. The size analysis for pure DOPC is presented in the SI.*

The SANS diffraction data is presented in Figure 4 for pure DOPC and DOPC mixed with 1% mass fraction of $dC_{18}$-PEO4 polymers dispersed in $D_2O$. The DOPC concentration is kept fixed at 5% mass fraction. The red solid lines represent the data modeling using the vesicle form factor as described in equations 1, 2 and 4. For both pure vesicle and for 1% $dC_{18}$-PEO4 polymer samples we illustrate data modeling using a ULV form factor ($N$ = 1). As shown in Figure 4, although we obtain a satisfactory description for the pure DOPC, we cannot explain the 1% $dC_{18}$-PEO4 data using a ULV model. We observe a $Q^{-3}$ power-law dependence over a $Q$-range, 0.02 to 0.08 Å$^{-1}$, which might be due to scattering from highly folded and convoluted surface arising from the adsorption of the polymer on the vesicle surface.[67, 68] Such surfaces are also visible in the cryo-TEM images (Figure 3 (a)) along with the formation of MLVs. In this case the presence of the polymer facilitates the MLV formation. We model the SANS data for the 1% $dC_{18}$-PEO4-DOPC sample using MLV model with the water core encapsulated by $N = 2$ lipid shells. The inset figure represents the SANS model used in data modeling. The parameters obtained from the TEM analysis, like the core radius, $R_c$, and the intermembrane water thickness, $t_w$ are used as initial estimation for fitting, to obtain the outer perimeter radius $R_p$ as presented in Table 2. Following DLS and cryo-TEM analysis we used a log-normal distribution for the polydispersity. From SANS we obtain a 9% increase in size of the DOPC vesicle from, $R_{SANS}$ = 53.6 ± 0.2 nm (0% dC18-PEO4) to 57.9 ± 0.5 nm (1% dC$_{18}$-PEO4). In presence of 1% dC18-PEO4 polymer, along with the formation of MLVs we observe a ≈ 5% reduction in the bilayer thickness, and development of a large intermembrane separation (~14 nm). From both SANS and cryo-TEM, an independent estimation yields a large polydispersity in $t_w$, possibly due to too many different intermembrane distances as suggested by TEM images (cf. supplementary information).

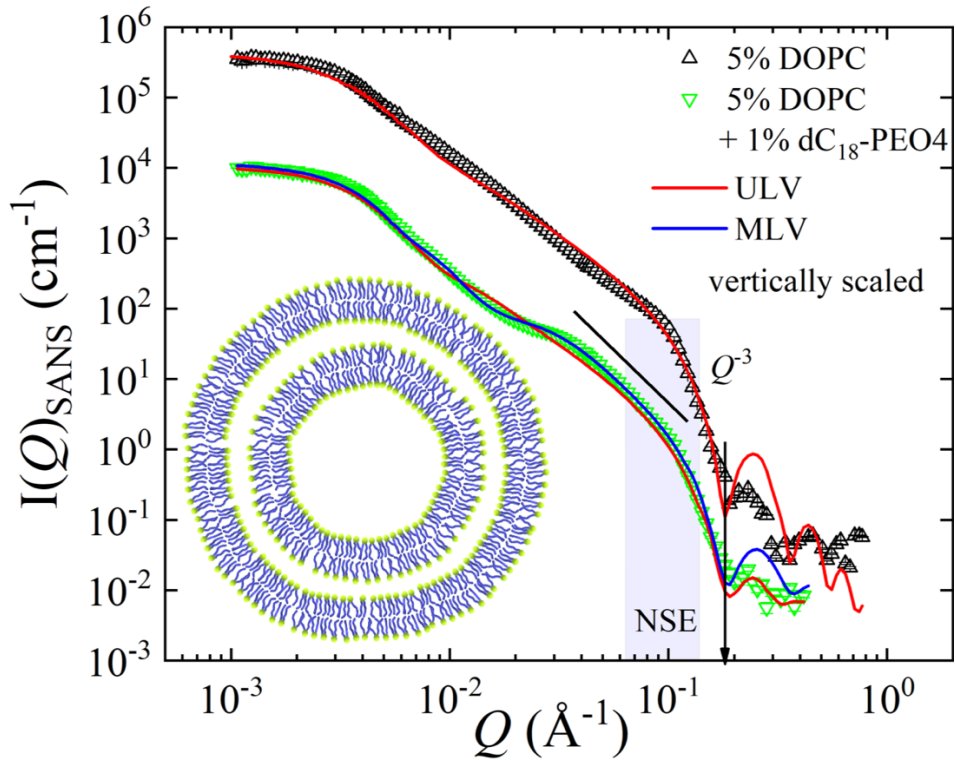

*Figure 4. SANS scattering data for pure DOPC and DOPC mixed with 1 wt% of dC18-PEO4 polymer dispersed in D$_2$O. The solid lines represent the ULV (N = 1) and MLV (N = 2) model*

*described in equations 1, 2 and 4. The shaded region depicts the Q-range over which NSE experiments are performed. Inset, depict the schematic illustration of the MLV (N = 2) model. The data are vertically scaled for proper visualization.*

*Table 2. Structural parameters from cryo-TEM and SANS. The number of layers, N, the outer perimeter radius of the vesicle, $R_p$, water core radius, $R_c$, intermembrane water thickness, $t_w$, bilayer thickness, $t_s$, and the corresponding size polydispersity. * Not visible in TEM images. **Not applicable to ULVs.*

| | Cryo-TEM | | SANS | |
|---|---|---|---|---|
| Samples | 0 % 1% dC$_{18}$-PEO4 | 1 % dC$_{18}$-PEO4 | 0 % dC$_{18}$-PEO4 | 1 % dC$_{18}$-PEO4 |
| $N$ | 1 | ~ 3 | 1 | 2 |
| $R_p$ (nm) | 51 ± 3 | 52 ± 4 | 54 ± 2 | 59 ± 2 |
| $R_c$ (nm) | 47 ± 2 | 28 ± 2 | 51 ± 2 | 37 ± 2 |
| $t_w$ (nm) | NA** | 14 ± 1 | NA** | 15 ± 1 |
| $t_s$ (nm) | NA* | NA* | 3.6 ± 0.1 | 3.4 ± 0.2 |
| $R_p$ Polydispersity (%) | 30 ± 7 | 34 ± 9 | 30 ± 2 | 40 ± 2 |
| $R_c$ Polydispersity (%) | 30 ± 6 | 52 ± 5 | 30 ± 2 | 40 ± 2 |
| $t_w$ Polydispersity (%) | NA** | 39 ± 4 | NA** | 41 ± 4 |

In Figure 5 we present the SAXS diffraction data for different concentrations of dC18-PEO4 polymers dispersed in 5% DOPC in D$_2$O. The data for DOPC without polymer is modeled using the bilayer structure (solid red lines) as described in equations 7, 8, and 9. The membrane bilayer lamellar structure for 0% dC$_{18}$-PEO4 sample can be described by $N = 2$ layers with a lamellar repeat distance $d = 6.6 ± 0.4$ nm, and $\delta_{HH} = 4.54 ± 0.01$ nm. The difference between N = 2 from SAXS with respect to the difference in N = 1 from SANS can be attributed to the heterogeneous nature of the sample. However, in presence of dC$_{18}$-PEO4 polymers the intensity at $Q ≈ 0.1$ Å$^{-1}$ (vertical dashed arrow) corresponding to the repetitive distance is vanishingly small. This effect of polymer induced modulation of the lipid bilayer can somewhat be captured by a very coarse-grained approach. We can see a resemblance of hard-sphere like packing, as explained by the solid black lines. An equivalent hard-sphere packing has been calculated from the product of hard-sphere form and structure factor given by equations 5-6. The analysis over 0.5 to 2.5% dC$_{18}$-PEO4 samples yields an average radius of $R_{HS}$ = 1.5 ± 0.05 nm which corresponds to the size of the hydrophobic moiety of the polymer. The corresponding hard sphere packing increases from $\phi_{HS}$ = 0.14 to 0.19 as the polymer concentration increased from 0.5 to 2.5%. This suggest an increase in packing of the hydrophobic moiety of the polymer into the lipid bilayer. The forward scattering pattern in presence of polymer is attributed to the difference in X-ray contrast between the hydrophobic alkyl group of the polymer and the lipid tail, $\Delta\rho$ = 2.7 × 10$^{-6}$ Å$^{-2}$. These calculations point towards the rapid entrapment of the hydrophobic alkyl group of the polymer into the lipophilic hydrocarbon core of the lipid bilayer. This phenomenon is illustrated schematically by the inset in Figure 5.

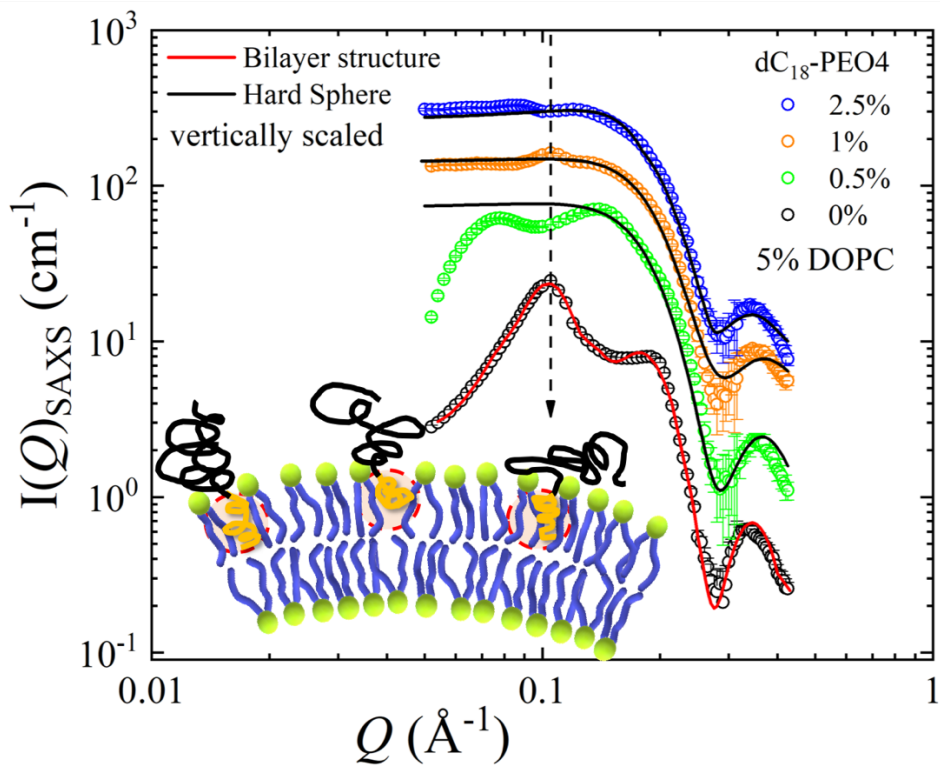

Figure 5. SAXS diffraction data of pure DOPC and DOPC mixed with varying mass fractions of dC18-PEO4 polymers dispersed in D₂O. The red and black solid lines are the data modeled using hard sphere and MLVs structure as described in equations 5 and 8 respectively. Inset, depict the *schematic illustration of the bilayer structure with the polymers wedged in. The data are vertically scaled for proper visualization.*

## 4.2  Dynamics

Figure 6 (a) illustrates the dynamic structure factor, $S(Q, t)$, measured by NSE for the 1% dC18-PEO4 -DOPC sample dispersed in $D_2O$, covering a $Q$-range from 0.063 to 0.139 Å⁻¹. The solid lines represent the model description using the ZG model as described in equation 12. The model accentuates deviations of the fit at low Fourier times while utilizing the ZG model which assumes the height-height correlations for membrane undulation but neglects other processes. The membrane rigidities, $\kappa_\eta / (k_B T)$ as calculated from equation 15 (for $N = 2$ layers) are listed in Table 3. For a better comparison, we included the result on pure DOPC ($N = 1$) from our previous studies.[37, 45]

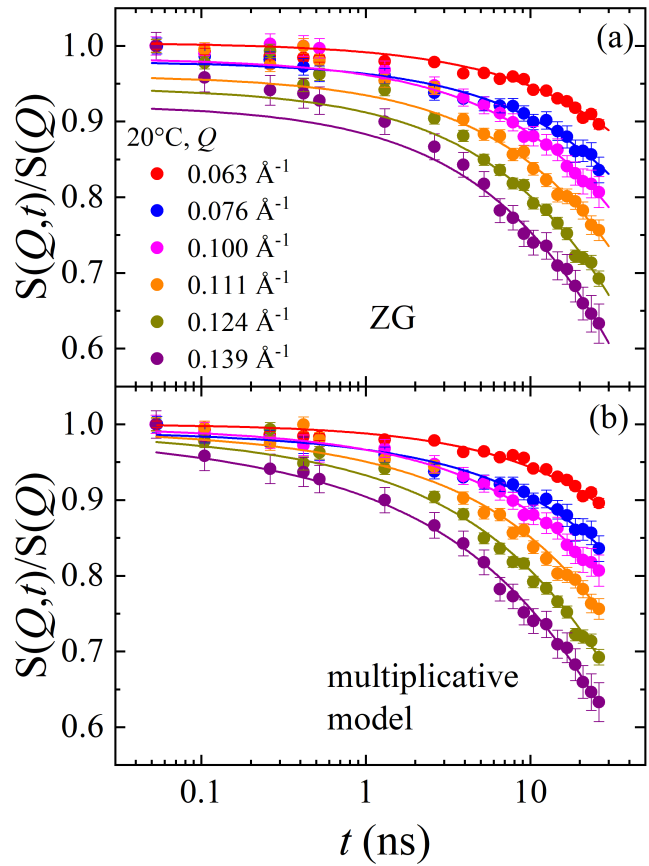

*Figure 6: Lin-log representations of the normalized dynamic structure factor, S(Q,t)/S(Q), as a function of Fourier time, t, for different Q's, DOPC mixed with 1% mass fraction of dC18-PEO4 polymer dispersed in $D_2O$ at 20 ℃. The same data sets are analyzed by fits using the (a) Zilman*

*Granek model (equation 12) and (b) the multiplicative model described in equation 11 that includes diffusion and confined motion. The error bars representing one standard deviation. A lin-lin plot is presented in the SI. The data for pure DOPC from our previous studies[37, 45] are presented in the SI.*

However, considering the fact that at least three different processes, the translational diffusion of the vesicle, the ZG membrane undulation and the contribution from tail motion of the lipids to $S(Q, t)$, have been identified in our previous publications, the NSE data can be modeled using the multiplicative model (equation 11) as illustrated in Figure 6 (b). The data modeling at lower Fourier times is related to the influence of lipid tail motion, which dominates at this time-scale. The obtained $\kappa_\eta/(k_B T)$ from equation 15 is reported in Table 3.

In order to examine the effect of different dynamics by a model independent approach we have calculated the MSD, $\langle \Delta r(t)^2 \rangle$, for the 1% dC18-PEO4 -DOPC sample using equation 17 along with the non-Gaussianity parameter, $\alpha_2$ as illustrated in Figure 7 (a) and (b), respectively. We have compared the results from pure DOPC.[37] Our earlier results from pure DOPC have a $t^{0.26\pm0.03}$ power-law dependence at low Fourier times, $t < 3$ ns, and the finite $\alpha_2$ seems to be related to the tail motion.[45] In this case the number of protons reflects the average scattering length density. However, in presence of deuterated 1% $dC_{18}$-PEO4 polymer we observe a $t^{0.30\pm0.05}$ power-law dependence where $\alpha_2 > 0$ only over a limited time window below $t \leq 0.42$ ns. This signifies that the presence of polymer restricts the fast lipid tail motion suggesting a damping mechanism. It should be noted that MSD is very important in biology and often plotted as a function of the temperature to understand thermal fluctuations and stability of biomacromolecules.[69]

The intermediate time range in Figure 7 (a) for pure DOPC shows a $t^{0.66}$ dependence for 3 ns < $t \leq 180$ ns, which can be attributed to membrane undulation as predicted by ZG model and a $t^{1.00\pm0.01}$ for $t > 180$ ns relates to the translational diffusion of the vesicle.[37, 45] In presence of 1% $dC_{18}$-PEO4 polymer the power law region proportional to $t^{0.66}$ extend over a broader range of times, 0.42 ns < $t \leq 10.5$ ns. Due to the observed suppression of the tail contribution at lower Fourier times, the contribution by the ZG could become better visible. This assumption would be consistent with the unchanged power law of $t^{0.66}$. As illustrated below, this extension can also be a consequence of the slightly increased bending modulus in case of a multilamellar vesicle. Within the time-range of the results of Figure 7a, the influence of diffusive motion is not visible in the data. Rather, we observe a weak $t^{0.51\pm0.02}$ power-law dependence for $t > 10.5$ ns as a result of strong encapsulation of the adsorbed polymeric layer on the membrane surface, which is supported by SANS (Figure 4).

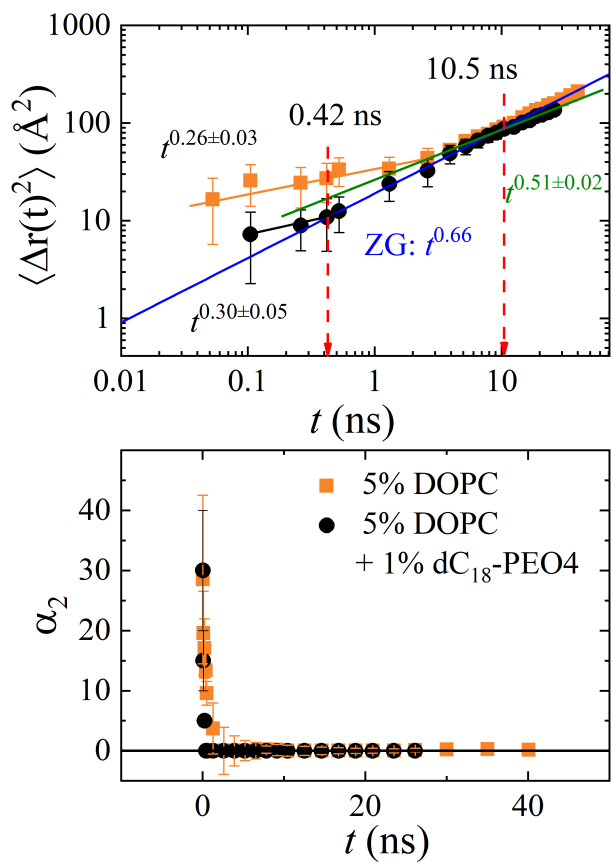

*Figure 7: (a) Mean square displacement, $\langle\Delta r(t)^2\rangle$, vs. Fourier time, t, for 5 wt% h-DOPC,[37] and 5 wt% of h-DOPC dispersed in 1 wt% of $dC_{18}$-PEO4 in $D_2O$ at 20 °C. The solid lines represent the experimental power-law dependence. (b) The corresponding non-Gaussian parameter $\alpha_2$.*

We can utilize the model free approach to obtain the changes in the membrane rigidity in presence of the polymer. By assuming that the ZG height-height correlations determine the dynamics in the full-time window of NSE, equation 18 can be used to calculate $\kappa_\eta/(k_BT)$ as a function of the Fourier time.[45] In Figure 8 the calculated $\kappa_\eta/k_BT$ for 1% $dC_{18}$-PEO4 -DOPC sample for MLVs with $N = 2$ is illustrated. An equivalent ULV membrane rigidity, $\kappa_\eta(ULV) =$ $\frac{\kappa_{\eta(MLV)}}{N}$ is used for comparison with 0% dC$_{18}$-PEO4 -DOPC sample (ULV) from our previous work.[37, 45] This approach magnifies any deviation from the ZG assumption. The shaded area in Figure 8 clearly elucidates a wider ZG regime (time independent behavior) in presence of 1% dC$_{18}$-PEO4 polymer.

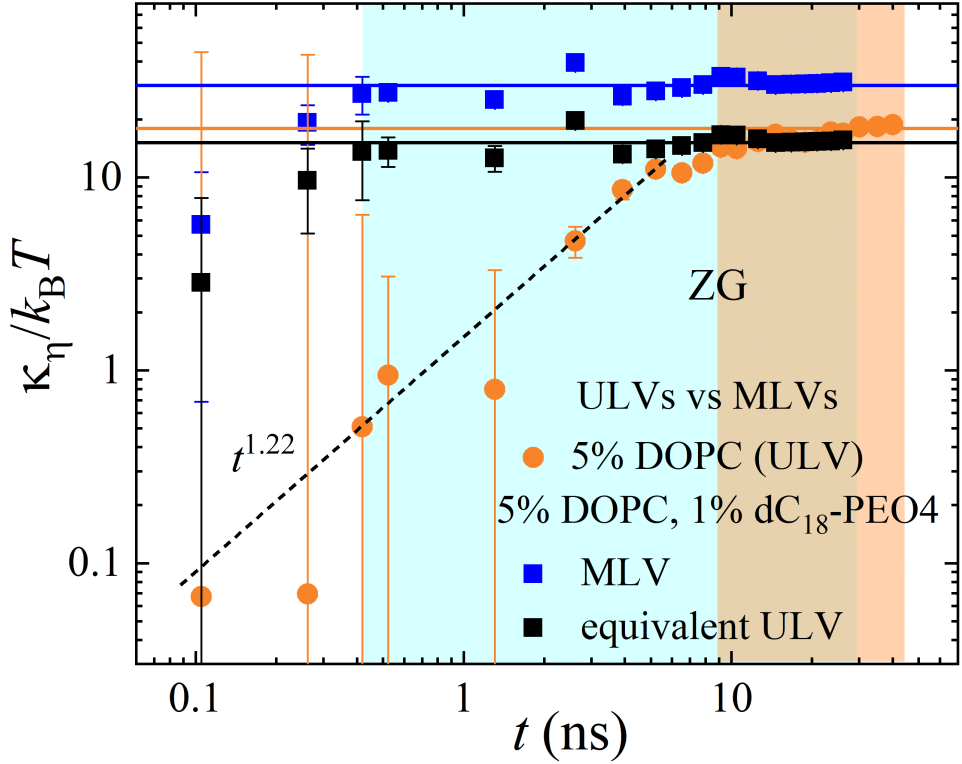

*Figure 8: The membrane rigidity, $\kappa_{\eta}$, divided by thermal energy, $k_B T$ , with the Boltzmann constant, $k_B$ , and the temperature, T, as a function of Fourier time. The data is calculated over the NSE time window from the MSD in Figure 7 (a), for 0 and 1 wt % dC$_{18}$-PEO4 polymer dispersed in D$_2$O at 20°C. The calculated average values from the flat ZG region is illustrated by the horizontal lines. These lines represent the bending modulus, $\kappa/k_B T$ and the values are listed in Table 1. The different power-laws are explained in the text.*

This observation is a consequence of equation 17. In the limit $Q \rightarrow 0$ we have $\kappa_{\eta}/k_B T \propto t^{2-3x}$. In this case, the ZG prediction with $x = 2/3 = 0.66$ yields a time independent behavior (solid lines), whereas the lipid tail motion has $x = 0.26$, yielding $t^{1.22}$ contribution (dotted line). This simple estimate assumes the effect of translational diffusion of the vesicle is negligible ($D_t = 0$), and $\alpha_2$

= 0. It should be noted, that for 1% $dC_{18}$-PEO4 sample the additional $t^{0.51}$ dependence in MSD analysis (Figure 7 (a)) for $t > 10.5$ ns has been eliminated by subtracting a $t^{0.47}$ contribution following the $\kappa_\eta/k_B T \propto t^{2-3x}$ dependence, with $x = 0.51$. This yield $\kappa_\eta/k_B T = 16 \pm 2$ for 1% dC18-PEO4 sample.

As the effect of diffusion on MSD is not visible for the 1% $dC_{18}$-PEO4 -DOPC sample (Figure 7 (a)), we have considered $D_t = 0$ in calculating $\kappa_\eta$ in Table 3. In Table 3 we have compared $\kappa_\eta/k_B T$ obtained from ZG (equation 12), multiplicative (equation 11) and MSD analysis (Figure 8) approaches using the $D_2O$ viscosity in equations 14 and 15. We have also included the calculation from ZG model (equation) using only the higher Fourier times (t > 5 ns) for the analysis, which results in $\kappa_\eta/k_B T$ values similar to those obtained by the multiplicative and MSD analysis. This emphasizes that the undulations prevail in the intermediate time range. For proper assessment of single bilayer rigidity of 1% $dC_{18}$-PEO4 -DOPC sample we have compared with 0% $dC_{18}$-PEO4 -DOPC sample by calculating the equivalent ULV membrane rigidity, $\kappa_\eta(N = 1) = \frac{\kappa_\eta(N=2)}{2}$. This analysis clearly elucidates the fact that in presence of 1% $dC_{18}$-PEO4 polymer the membrane rigidity in each bilayer decreases.

*Table 3. Membrane rigidity $\kappa_\eta$ obtained for 1% $dC_{18}$-PEO4 -DOPC sample using different models for $D_t = 0$, $\eta = \eta_{D2O}$. We have used results of 0% $dC_{18}$-PEO4 -DOPC sample from our previous study.[37, 45, 56] For comparison with pure DOPC we have included $\kappa_\eta$ for N =2 and N = 1 layers.*

| $\kappa_\eta/(k_B T)$ | | | | | | |
|---|---|---|---|---|---|---|
| Parameters | Concentration dC$_{18}$-PEO4, wt% | N | ZG Analysis (Full time range) (equation 12) | ZG Analysis (t >5 ns) | Multiplicative model (equation 11) | MSD analysis |
| $D_t = 0$ $\eta = \eta_{D2O}$ | 0 | 1 | 26 ± 1 | 20 ± 2 | 21 ± 2 | 18 ± 2 |
| | 1 | 2 | 26 ± 5 | 28 ± 5 | 29 ± 5 | 30 ± 3 |
| | 1 | 1 | 13 ± 3 | 14 ± 4 | 15 ± 3 | 15 ± 2 |

Bringing together the information from SAXS and SANS we can illustrate the effect of adding d-S100 polymer from outside the vesicles in aqueous solution in Figure 9. The polymer forms micelles in aqueous solution as shown by Zinn et al.[61] We observed from SAXS analysis, that in presence of lipid vesicles the unimers from the micelles consisting of hydrophobic alkyl and hydrophilic PEO segments, prefers to wedge into the bilayer membrane. The hydrophobic alkyl group of the polymer resides in the lipophilic hydrocarbon core of the lipid bilayer. Such preferential migration of the unimers into the membrane was also observed in case of bicontinuous microemulsions.[70] This phenomenon causes structural defects in the lipid bilayers, as illustrated in Figure 9 a, supported by the TEM images. This disruption provides an opportunity for bridge formation (Figure 9 b) between individual vesicles that leads to transport of free lipids, that acts as a nucleation site for the formation of a new bilayer, resulting in formation of MLVs (Figure 9 c) determined by SANS. The partial disruption of the bilayer by the polymer is responsible for reduction in each bilayer membrane rigidity, determined by NSE. The trapped unimers in the bilayer have their long hydrophilic PEO chains dangling outside the vesicle interface into the water. They exert a hydrophilic tension on the lipid membrane causing an overall expansion in size of the vesicles, verified by DLS and SANS. The simultaneous formation of MLVs contributes further to the increase in the size of the vesicles.

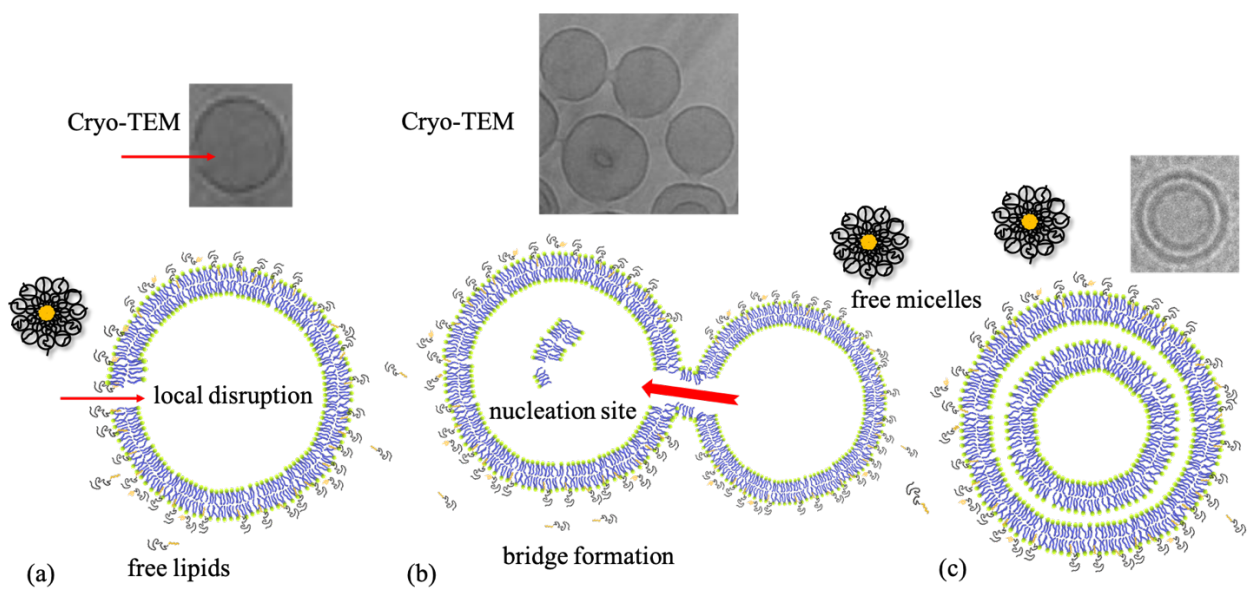

*Figure 9: Schematic illustration of the formation of (a) local disruption, (b) bridge formation, and (c) multilayer vesicles (MLVs) in presence of dC$_{18}$-PEO4 polymers.*

## 5   CONCLUSION

We performed a detailed analysis of the vesicles structure, morphology, and dynamics of vesicles in presence of $dC_{18}$-PEO4 polymers using Cryo-TEM, DLS, SAXS, SANS and NSE experiments. Our results confirmed the existence of multilayer vesicles and an increase in size of the vesicles in presence of the polymers. We demonstrated that polymers are trapped in the lipid bilayer, causing a partial disruption in the vesicle. This phenomenon is attributed to the reduction in bending rigidity per unit bilayer. The MSD analysis indicates the reduction of trapped motion of the lipid tail over a broad length and time scale accessible by NSE. Our results emphasized the opportunities of using unique hydrophilic-hydrophobic di-block polymers that can disrupt membrane and control their dynamics with possible application in topical drug or nutraceutical formulations.

## ACKNOWLEDGEMENT

The neutron scattering work is supported by the U.S. Department of Energy (DOE) under EPSCoR Grant No. DE-SC0012432 with additional support from the Louisiana Board of Regents. Access to the neutron spin echo spectrometer and small-angle scattering instruments was provided by the Center for High Resolution Neutron Scattering, a partnership between the National Institute of Standards and Technology and the National Science Foundation under Agreement No. DMR-1508249. Research conducted at the Spallation Neutron Source (SNS) at Oak Ridge National Laboratory (ORNL) was sponsored by the Scientific User Facilities Division, Office of Basic Energy Sciences, U.S. DOE. We thank Lin Yang and Shirish Chodankar from 16-ID, LIX beamline at National Synchrotron Light Source (NSLS) II. The LiX beamline is part of the Life Science Biomedical Technology Research resource, primarily supported by the National Institute of Health, National Institute of General Medical Sciences under Grant P41 GM111244, and by the Department of Energy Office of Biological and Environmental Research under Grant KP1605010, with additional support from NIH Grant S10 OD012331. As a NSLS II facility resource at Brookhaven National Laboratory, work performed at Life Science and Biomedical Technology Research is supported in part by the US Department of Energy, Office of Basic Energy Sciences Program under Contract DE-SC0012704.

## DISCLAIMER

Certain trade names and company products are identified in order to specify adequately the experimental procedure. In no case does such identification imply recommendation or